\documentclass{article}
  
\usepackage[utf8]{inputenc}
\usepackage[T1]{fontenc}
\usepackage[mathscr]{euscript}
\usepackage{geometry}
\usepackage{amssymb}
\usepackage{amsmath}
\usepackage{graphicx}
\usepackage{cite}
\title{A no-go theorem for regular black holes}
\author{Stefano Chinaglia\footnote{e-mail: s.chinaglia@unitn.it} \\
\smallskip \\
\it Dipartimento di Fisica, Università di Trento \\
\it Via Sommarive 14, 38123 Trento, Italia}
\date{\today}

\begin{document}
\maketitle

\begin{abstract}

In this article we discuss a no-go theorem for generating regular black holes from a Lagrangian theory. We prove that the general solution has always a Schwarzschild-like term $c/r$, as long as the matter Lagrangian depends neither on the metric, nor on its derivatives; we also prove that, under suitable additional conditions, these two conditions are also equivalent to $g_{00}g_{11} = -1$. Finally, we prove that $c/r$ is the only non-Lagrangian singularity eventually present into the solution.

\end{abstract}

\section{Introduction}

An important branch of the research in black hole physics is the search for regular black holes (RBH). It is indeed well known that many solutions of the Einstin equations (EE) are actually singular, e.g. Schwarzschild or Reissner-Nordström. However nature is expected to be singularity-free, so such solutions indicate either a breakout or an incompleteness of their generating theory.

In principle, one may circumnavigate the problem, supposing that black holes do not exist, as objects having an horizon (so that they also do not have any central singularity) and they are just {\it quasi} black holes, i.e. compact objects, which are however slightly larger than their horizon itself: examples of such objects may be found in \cite{Majumdar47, Papapetrou47, Mazur01, Lemos07}. And actually the concept itself of RBH is open to some criticisms, such as the instability of the inner horizon \cite{Maeda05} (for a very recent argument against the standard notion of RBH, see \cite{Visser18}). However, the recent observations by the LIGO-VIRGO collaboration \cite{LIGO1, LIGO2, LIGO3, LIGO4, LIGO5, LIGO6, VIRGO} suggest that black holes actually exist and consequently they are expected not to be singular objects. It is worth to notice that such observation do not exclude the possibility of quasi-black holes, but it does not seem to be favoured.

Assuming then that black holes actually exist, it is importat to see how they can be regularized. One of the very first attempts to write a regular metric was done by Y.S. Duan in 1954 \cite{Duan}, but his worked remained ignored for a long time. The first example of RBH metric was proposed by J. Bardeen during the GR5 conference fourteen years later \cite{Bardeen}: since then, many other examples have been provided, such as \cite{Israel88, Poisson89, Hayward06, Cognola13, Dymnikova15, Horava09, Bronnikov01, Dymnikova04, Dymnikova92, Culetu15, Pradhan15, Horava09_1, Kehagias09, Cai10, Elizalde02, Nicolini06, Ansoldi07, Modesto10, Maeda15, Ma15, Johannsen13, Rodrigues16, Fan16, Beato98, Frolov_1, Frolov_2, Frolov_3, Frolov_4, Frolov_5, Frolov_6, Frolov_7, Cisterna12, Vagenas14, Vagenas14_1, Giugno17, Contreras17, Odintsov17_1}. For some review on the topic, see e.g \cite{Ansoldi08, Spallucci17, Colleaux17}.

Each of these solution has an asymptotic Schwarzschild behaviour at infinity and exactly reduces to Schwarzschild when the deformation parameters are turned off (a cosmological term $\propto r^2$ is sometimes present, but it can be easily managed). On the contrary, near the origin solutions have a de Sitter core, which is known as Sakharov criterion \cite{Sakharov}. We should however say that the fulfillment of the Sakharov criterion from one solution $f$ of a given Lagrangian $\mathscr{L}$ does not guarantee that every solutions generated by it will behave in this way: as we will see, indeed, this is not the case for a number of models.

The point is indeed that one should check if the Lagrangian is not such that the singularity is in some way allowed to come back. This is precisely what happens e.g. in the so called Non-linear ElectroDynamics (NED), which tries to avoid the GR singularity making use of some deformed electrodynamical Lagrangian. What turns out in those cases is that, if $f$ is a solution of the EE, then also $f + c/r$ is, where $c$ is a completely arbitrary integration constant: and there is no way to avoid it.

The aim of the present work is to build the most general criterion a Lagrangian or a solution should satisfy, in order to avoid the awakening of such singular Schwarzschild-like terms. A first version of this no-go theorem has been presented in \cite{Chinaglia18}. In that work, we already studied the case of a theory minimally coupled to GR, i.e. $\mathscr{L} = R - \mathscr{L}_{matter}$. In this work, we are going to generalize our previous argument to the most possible general theory, making no assumption at all on the Lagrangian.

The paper is organized as follows. In sec. 2 we review an example of problematic theory, i.e. the NED approach; in sec. 3 we present the assumptions and the statemant of our no-go theorem, while in sec. 4 we will present and prove a preparatory lemma. Sec. 5 and sec. 6 are devoted to the proof of the theorem iteself and in sec. 7 we present a corollary. Finally, in sec. 8 we will have some final remark.

\section{The NED approach}

In this section, we want to summerize the so called NED approach: we do so, in order to show an example of model, whose Lagrangian admit the Schwarzschild-like term. This approach first appeared in \cite{Beato98} by A. Garcia and E. Ayon-Beato, while other paper on this topic are e.g. \cite{Dymnikova04, Dymnikova92, Beato00, Salazar87, Breton03, Novello00, Dymnikova15, Culetu15, Ma15, Dymnikova04_1, Dymnikova05, Dymnikova03, Junior15, Toshmatov18}. Its aim is to regularize the gravitational singularity, coupling an electromagnetic potential to gravity:

\begin{equation}
\label{A1}
\mathscr{I} = \int d^4 x \ \sqrt{-[g]} \left( \frac{R}{2} - 4\pi \mathscr{L}(I) \right) \ ,
\end{equation}

where $R$ is the Ricci scalar, $I = \frac{1}{4} F^{\mu\nu} F_{\mu\nu}$ is an electromagnetic-like tensor and $\mathscr{L}(I)$ is a suitable function of it. The matter Lagrangian $\mathscr{L}$ is chosen so that it reduces asymptotically to the standard Maxwell Lagrangian $\propto I$. Since the coupling is minimale and the Lagrangian only depends on $I$, the field equations read

\begin{equation}
\label{A2}
G^\nu_\mu = -8\pi \left( F_{\alpha\mu} \partial_I \mathscr{L} F^{\nu\alpha} + \mathscr{L}\delta^\nu_\mu \right) \ ,
\end{equation}

\begin{equation}
\label{A3}
\nabla^\mu (F_{\mu\nu} \partial_I \mathscr{L}) = 0 \ .
\end{equation}

Eq. (\ref{A2}) is the EE, while eq. (\ref{A3}) is the modified Maxwell equation. In this brief review, we follow \cite{Chinaglia17_2}, however we should note that there exists also an equivalent approach, i.e. the so called dual $P$ approach \cite{Salazar87}, for which some discussions turn to be simpler.

In the following, we will consider a spacetime in the form

\begin{equation}
\label{A4}
ds^2 = -f(r)dt^2 + \frac{1}{f(r)} dr^2 + r^2 d\Omega^2 \ .
\end{equation}

The equivalence $f=g$ can be proved in the dual $P$ approach, as shown in \cite{Chinaglia15}. If so, the field equation (\ref{A2})-(\ref{A3}) read

\begin{equation}
\label{A5}
\frac{f'r + f - 1}{r^2} = 8\pi \left( -2 I\partial_I \mathscr{L} + \mathscr{L} \right) = -8\pi \rho \ ,
\end{equation}

\begin{equation}
\label{A6}
\partial_r \left( r^2 \partial_I \mathscr{L} F^{0r} \right) = 0 \ .
\end{equation}

These equations are enough for our purposes, so we won't explore deeper the NED approach. We remand the interested reader to the original material or to the review \cite{Colleaux17}.

The key point of eq. (\ref{A5})-(\ref{A6}) is that one can prove, under the staticity assumption, that $I = - \frac{1}{2} F_{0r}^2$, with no explicit dependence on the metric or its derivatives: this makes also the Lagrangian, and thus the density, independent on the metric or its derivatives. This leads directly to the catastrophe we already mentioned in the introduction: if we assume $f$ being a solution and we insert $f + c/r$ into eq. (\ref{A5}), we have

\begin{equation}
\label{A7}
\frac{f'r - \frac{c}{r^2} r + f + \frac{c}{r} - 1}{r^2} = \frac{f'r + f - 1}{r^2} = -8\pi \rho \ .
\end{equation}

Since $\rho$ does not depend the metric, it has exactly the same shape of eq. (\ref{A5}), so that also $f + c/r$ is still a solution; and since $c$ is an integration constant, there is no way to fix it.

By the way, this result should not surprise, since eq. (\ref{A5}) is a linear non-homogeneous ordinary differential equation: its solution is then given by a particular solution of the non-homogeneous equation plus the general solution of the homogeneous one; and the homogeneous solution is just the Schwarzschild term $c/r$.

In order to show that $c$ cannot be fixed, even considering the asymptotic behaviour, fix on a specific metric, e.g. the Bardeen solution \cite{Bardeen}

\begin{equation}
\label{A8}
f(r) = 1 - \frac{2K r^2}{(r^2 + l^2)^{3/2}} \ ,
\end{equation}

where $K$ is an integration constant. It sounds natural to fix $K$ as the mass of the black hole, but this is true only in the case of $c=0$: indeed, the mass is half of the coefficient of the $1/r$ term in the asymptotic expansion. Thus, adding to solution (\ref{A8}) the $c/r$ term, its asymptotic behaviour becomes

\begin{equation}
\label{A9}
f(r \rightarrow \infty) = 1 - \frac{2K-c}{r} + o(r^{-2}) \ .
\end{equation}

It is then clear that the mass is now $m = K - c/2$. Since both $K$ and $c$ are arbitrary constants and since we have a constraint only on their sum (the only asymptotic requirement is $2K-c = 2m$), we are not free to fix any of them. $c=0$ still remains an admissible choice, but nothing guarantees it should be performed (unless some physical criterion is found, as done in \cite{Chinaglia17}).

\section{Assumptions and statement}

We start with the usual metric of spherical symmetry:

\begin{equation}
\label{B1}
ds^2 = - f(r) dt^2 + \frac{1}{g(r)} dr^2 + r^2 d\Omega^2 \ ,
\end{equation}

where $d\Omega^2$ is the surface element of a sphere of unitary radius. Changes of topology do not seem to modify the general argument, but we won't discuss them here. We now write the action in the form

\begin{equation}
\label{B2}
\mathscr{I} = \int d^4 x \sqrt{-[g]} \mathscr{L} \ ,
\end{equation}

where $[g]$ is the determinant of the metric and $\mathscr{L}$ is the Lagrangian density. This Lagrangian is the most general possible one, containing any gravity and matter term, coupled in the most possible general way. In $\mathscr{L}$ may appear a general number of any matter field (we will deal with a single one, just for simplicity) and a general number of any scalar invariant, such as $R$, $R_{\mu\nu}R^{\mu\nu}$, $R_{\mu\nu\rho\sigma}R^{\mu\nu\rho\sigma}$ etc. without any limitation of mixing. For a mere reason of convenience, however, we write action (\ref{B2}) as

\begin{equation}
\label{B3}
\mathscr{I} = \int d^4 x \sqrt{-[g]} \left( R - \mathscr{L}_{eff} \right) \ ,
\end{equation}

where $R$ is the Ricci scalar and $\mathscr{L}_{eff} \equiv R - \mathscr{L}$. The effective Lagrangian plays now the role of effective matter Lagrangian, minimally coupled to gravity. We have always the right to take this interpretation, because it is just a readjustment in the factors.

Before proceeding, we should recall that {\it any} Lagrangian can be written in the form (\ref{B3}) and this gives enough generality to our approach. Now, using the variational principle, we find the following field equations:

\begin{equation}
\label{B4}
R_{\mu}^{\nu} - \frac{1}{2} R \delta_{\mu}^{\nu} = g^{\alpha\nu} \frac{\partial \mathscr{L}_{eff}}{\partial g^{\mu\alpha}} - \frac{1}{2} \mathscr{L}_{eff} \delta_{\mu}^{\nu} - g^{\alpha\nu} \frac{\partial}{\partial x^{\rho}} \frac{\partial \mathscr{L}_{eff}}{\partial \frac{\partial g^{\mu\alpha}}{\partial x^{\rho}}} \ ,
\end{equation}

\begin{equation}
\label{B5}
\nabla^{\mu} \frac{\delta \mathscr{L}_{eff}}{\delta \nabla^{\mu} \phi} - \frac{\delta \mathscr{L}_{eff}}{\partial \phi} = 0 \ .
\end{equation}

Eq. (\ref{B4}) collectively summarizes the EE, while eq. (\ref{B5}) is the Lagrange equation for the matter field $\phi$ (here we did not show its indices, which can be kept quit general). It is worth to notice that in $\mathscr{L}_{eff}$, $R$ does not depend on the matter field, so that eq. (\ref{B5}) also reads

\begin{equation}
\label{B6}
\nabla^{\mu} \frac{\delta \mathscr{L}}{\delta \nabla^{\mu} \phi} - \frac{\delta \mathscr{L}}{\partial \phi} = 0 \ ,
\end{equation}

where $\mathscr{L}$ is once again the true Lagrangian; if the decoupling is possible, it reduces to the standard matter Lagrangian (as in the case of minimal coupling). Finally, we assume that the pair $(\bar{f}, \bar{g})$ solves the field equation and we define $f \equiv \bar{f} + c/r$ and $g$ similarly (with the same integration constant $c$).

If so, we are finally able to present the statement of the theorem:

\bigskip

\begin{minipage}{.8\textwidth}

{\it The following two statements are equivalent:}

\begin{equation}
\begin{split}
\label{B7}
&\text{(a)} \ (f, g) \ \text{\it is a solution;} \\
&\text{(b)} \ \partial_{(\bar{f}, \bar{g})} \mathscr{L} = 0 \ \text{and} \ \partial_{(\bar{f}', \bar{g}')} \mathscr{L} = 0 \ .
\end{split}
\end{equation}

\end{minipage}

\bigskip

Moreover, if we assume also the subsidiary hypothesis that $\partial_{(\bar{f}, \bar{g})} \mathscr{L} = 0$ or $\partial_{(\bar{f}', \bar{g}')} \mathscr{L} = 0$ (one is enough), the statement reads

\bigskip

\begin{minipage}{.8\textwidth}

{\it The following three statements are equivalent:}

\begin{equation}
\begin{split}
\label{B8}
&\text{(a)} \ (f, g) \ \text{\it is a solution;} \\
&\text{(b)} \ \partial_{(\bar{f}, \bar{g})} \mathscr{L} = 0 \ \text{and} \ \partial_{(\bar{f}', \bar{g}')} \mathscr{L} = 0 \ ; \\
&\text{(c)} \ \bar{f} = \bar{g} \ .
\end{split}
\end{equation}

\end{minipage}

\bigskip

This second statement reduces to the case we already considered in \cite{Chinaglia18}. It is also worth to notice that this second statement holds only if we assume the extra hypothesis $\partial_{(\bar{f}, \bar{g})} \mathscr{L} = 0$ or $\partial_{(\bar{f}', \bar{g}')} \mathscr{L} = 0$, because otherwise we are able to prove that it does not hold. There are actually some examples of it, such as the BLZ black hole \cite{Balakin16, Balakin16_1}, which satisfies (c), but not (a) and (b).

\section{Lemma}

The general statement of the lemma reads

\begin{equation}
\label{C1}
\partial_{(\bar{f}, \bar{g})} \mathscr{L} = 0 \ \text{and} \ \partial_{(\bar{f}', \bar{g}')} \mathscr{L} = 0 \qquad \Leftrightarrow \qquad \partial_{(\bar{f}, \bar{g})} T_{\mu}^{\nu} = 0 \ \text{and} \ \partial_{(\bar{f}', \bar{g}')} T_{\mu}^{\nu} = 0 \ .
\end{equation}

This means that the SET does not depend on the metric and its derivatives, if and only if the effective Lagrangian does not depend on the metric and its derivatives. Notice that both conditions are necessary for the double implication.

\subsection{$\partial_{(\bar{f}, \bar{g})} \mathscr{L} = 0 \ \text{and} \ \partial_{(\bar{f}', \bar{g}')} \mathscr{L} = 0 \qquad \Rightarrow \qquad \partial_{(\bar{f}, \bar{g})} T_{\mu}^{\nu} = 0 \ \text{and} \ \partial_{(\bar{f}', \bar{g}')} T_{\mu}^{\nu} = 0$}

We start computing directly $\partial_{\bar{g}} T_{\mu}^{\nu}$: we have

\begin{equation}
\label{C2}
\begin{split}
\partial_{\bar{g}} T_{\mu}^{\nu} &= \partial_{\bar{g}} \left( g^{\alpha\nu} \frac{\partial \mathscr{L}_{eff}}{\partial g^{\mu\alpha}} - \frac{1}{2} \mathscr{L}_{eff} \delta_{\mu}^{\nu} - g^{\alpha\nu} \frac{\partial}{\partial x^{\rho}} \frac{\partial \mathscr{L}_{eff}}{ \frac{\partial g^{\mu\alpha}}{\partial x^{\rho}} } \right) \\
&= \partial_{\bar{g}} g^{\alpha\nu} \frac{\partial \mathscr{L}_{eff}}{\partial g^{\mu\alpha}} + g^{\alpha\nu} \partial_{\bar{g}} \frac{\partial \mathscr{L}_{eff}}{\partial g^{\mu\alpha}} - \frac{1}{2} \partial_{\bar{g}} \mathscr{L}_{eff} \delta_{\mu}^{\nu} - \partial_{\bar{g}} g^{\alpha\nu} \frac{\partial}{\partial x^{\rho}} \frac{\partial \mathscr{L}_{eff}}{ \frac{\partial g^{\mu\alpha}}{\partial x^{\rho}} } - g^{\alpha\nu} \partial_{\bar{g}} \frac{\partial}{\partial x^{\rho}} \frac{\partial \mathscr{L}_{eff}}{ \frac{\partial g^{\mu\alpha}}{\partial x^{\rho}} } \\
&= g^{\alpha\nu} \partial_{\bar{g}} \frac{\partial \mathscr{L}_{eff}}{\partial g^{\mu\alpha}} + \left( \delta_{\mu}^1 \delta_1^{\nu} - \frac{1}{2} \delta_{\mu}^{\nu} \right) \partial_{\bar{g}} \mathscr{L}_{eff} - \delta_{\mu}^1 \delta_1^{\nu} \frac{\partial}{\partial r} \frac{\partial \mathscr{L}_{eff}}{\partial g'} - g^{\alpha\nu} \partial_{\bar{g}} \frac{\partial}{\partial x^{\rho}} \frac{\partial \mathscr{L}_{eff}}{ \frac{\partial g^{\mu\alpha}}{\partial x^{\rho}} } \\
&= 0
\end{split}
\end{equation}

Once again by direct computation, we study now $\partial_{\bar{g}'} T_{\mu}^{\nu}$: we have

\begin{equation}
\label{C3}
\begin{split}
\partial_{\bar{\bar{g}'}} T_{\mu}^{\nu} &= \partial_{\bar{g}'} \left( g^{\alpha\nu} \frac{\partial \mathscr{L}_{eff}}{\partial g^{\mu\alpha}} - \frac{1}{2} \mathscr{L}_{eff} \delta_{\mu}^{\nu} - g^{\alpha\nu} \frac{\partial}{\partial x^{\rho}} \frac{\partial \mathscr{L}_{eff}}{ \frac{\partial g^{\mu\alpha}}{\partial x^{\rho}} } \right) \\
&= g^{\alpha\nu} \partial_{\bar{g}'} \frac{\partial \mathscr{L}_{eff}}{\partial g^{\mu\alpha}} - \frac{1}{2} \partial_{\bar{g}'} \mathscr{L}_{eff} \delta_{\mu}^{\nu} - g^{\alpha\nu} \partial_{\bar{g}'} \frac{\partial}{\partial x^{\rho}} \frac{\partial \mathscr{L}_{eff}}{ \frac{\partial g^{\mu\alpha}}{\partial x^{\rho}} } \\
&= 0
\end{split}
\end{equation}

The proof for $\partial_{\bar{f}} T_{\mu}^{\nu}$ and $\partial_{\bar{f}'} T_{\mu}^{\nu}$ is conducted similarly.

\subsection{$\partial_{(\bar{f}, \bar{g})} \mathscr{L} = 0 \ \text{and} \ \partial_{(\bar{f}', \bar{g}')} \mathscr{L} = 0 \qquad \Leftarrow \qquad \partial_{(\bar{f}, \bar{g})} T_{\mu}^{\nu} = 0 \ \text{and} \ \partial_{(\bar{f}', \bar{g}')} T_{\mu}^{\nu} = 0$}

This proof is slightly more difficult. We start studying the derivative with respect to $g$. Since the SET does not depend on the metric, i.e. $\partial_{\bar{g}} T_{\mu}^{\nu} = 0$, we have

\begin{equation}
\label{C4}
\partial_{\bar{g}} g^{\alpha\nu} \frac{\partial \mathscr{L}_{eff}}{\partial g^{\mu\alpha}} + g^{\alpha\nu} \partial_{\bar{g}} \frac{\partial \mathscr{L}_{eff}}{\partial g^{\mu\alpha}} - \frac{1}{2} \partial_{\bar{g}} \mathscr{L}_{eff} \delta_{\mu}^{\nu} - \partial_{\bar{g}} g^{\alpha\nu} \frac{\partial}{\partial x^{\rho}} \frac{\partial \mathscr{L}_{eff}}{ \frac{\partial g^{\mu\alpha}}{\partial x^{\rho}} } - g^{\alpha\nu} \partial_{\bar{g}} \frac{\partial}{\partial x^{\rho}} \frac{\partial \mathscr{L}_{eff}}{ \frac{\partial g^{\mu\alpha}}{\partial x^{\rho}} } = 0 \ .
\end{equation}

Since this equations holds for any choice of the indices, we impose $\mu = \nu = 3$: thus eq. (\ref{C4}) reduces to

\begin{equation}
\label{C5}
- \frac{1}{2} \partial_{\bar{g}} \mathscr{L}_{eff} = 0 \ .
\end{equation}

Indeed, the request of spherical symmetry forbids the Lagrangian to depend on the angular variables, so that it does not depend on $g_{33}$ (unless in trivial ways, such as $g_{33}g^{33}$). This immediately closes the first part of the proof. Re-inserting this into eq. (\ref{C4}), we have a second condition on the Lagrangian, i.e.

\begin{equation}
\label{C6}
- \delta_{\mu}^1 \delta_1^{\nu} \frac{\partial}{\partial r} \partial_{\bar{g}'} \mathscr{L}_{eff} = 0 \ ,
\end{equation}

meaning that either the Lagrangian does not depend on $\bar{g}'$, or that $\mathscr{L}_{eff} = A(r) + C \bar{g}'$, where $A(r)$ does not depend on $\bar{g}'$, while $C$ is just a constant. Since a similar proof holds also for $f$, this means that

\begin{equation}
\label{C7}
\mathscr{L}_{eff} = A(r) + B \bar{f}'+ C \bar{g}' \ .
\end{equation}

Eq. (\ref{C7}) is howver unacceptable, unless $B=C=0$: this because it violates the assumption that the SET deos not depend on the metric or its derivatives. Indeed, computing $T_0^0$, we have

\begin{equation}
\label{C8}
T_0^0 = \frac{3}{2} B \bar{f}' - \frac{1}{2} C \bar{g}' - \frac{1}{2} A \ .
\end{equation}

Since $A$ depends only on $r$, we should impose $B=C=0$; and this is enough to end the proof.

Before closing this section and switching to the proof of the theorem, we should note that the lemma has also a reduced version:

\begin{equation}
\label{C9}
\partial_{(\bar{f}', \bar{g}')} \mathscr{L} = 0 \qquad \Leftrightarrow \qquad \partial_{(\bar{f}', \bar{g}')} T_{\mu}^{\nu} = 0 \ ,
\end{equation}

i.e. the Lagrangian does not depend on the derivatives of the metric (but may depend on the metric itself) if and only if the SET does not. It can be easily proved using the previous equations. The contrary (i.e. $\partial_{(\bar{f}, \bar{g})} \mathscr{L} = 0 \Leftrightarrow \partial_{(\bar{f}, \bar{g})} T_{\mu}^{\nu} = 0$) is in general not true.

\section{Proof of the theorem $-$ part 1}

We start proving statement (\ref{B7}). Later on, we will make the extra assumption and prove the remaining part of statement (\ref{B8}).

\subsection{ $(f, g) \ \text{is a solution} \qquad \Rightarrow \qquad \partial_{(\bar{f}, \bar{g})} \mathscr{L} = 0 \ \text{and} \ \partial_{(\bar{f}', \bar{g}')} \mathscr{L} = 0$}

We start writing down the 00 component of the EE, both for $(\bar{f}, \bar{g})$ and $(f, g)$;, we have

\begin{equation}
\label{D1}
\frac{\bar{g}' r + \bar{g} -1}{r^2} = \bar{T_0^0} \ ,
\end{equation}

\begin{equation}
\label{D2}
\frac{g'r + g -1}{r^2} = \frac{\bar{g}' r + \bar{g} -1}{r^2} = T_0^0 \ .
\end{equation}

Subtracting the two equations side to side, we immediately find that the 00 component of the SET does not depend on the solution itself: $T_0^0 (f, g) - T_0^0 (\bar{f}, \bar{g})$.

This does not mean that we can apply the lemma, since still we don't have any information on the remaining independent components of the SET. However we are still allowed to say that the $\partial_{(f, g)} T_0^0 = 0$, i.e. respectively

\begin{equation}
\label{D3}
f \partial_g \partial_f \mathscr{L}_{eff} - \frac{1}{2} \partial_g \mathscr{L}_{eff} - 2f' \partial_g \partial_{f'} \mathscr{L}_{eff} - f \partial_g \frac{\partial}{\partial r} \partial_{f'} \mathscr{L}_{eff} = 0 \ ,
\end{equation}

\begin{equation}
\label{D4}
f \partial_f^2 \mathscr{L}_{eff} + \frac{3}{2} \partial_f \mathscr{L}_{eff} - 2f' \partial_f \partial_{f'} \mathscr{L}_{eff} - f \partial_f \frac{\partial}{\partial r} \partial_{f'} \mathscr{L}_{eff} = 0 \ ,
\end{equation}

If $\partial_{f'} \mathscr{L}_{eff} = 0$, the system can be solved explicitly and from it, it is possible to close the proof. This is what we did in \cite{Chinaglia18}. Here things are more difficult and the system cannot be solved analytically; however, we don't need an explicit solution, but we only need to show that assuming either $\partial_{(\bar{f}, \bar{g})} \mathscr{L}_{eff} = 0$ or $\partial_{(\bar{f}', \bar{g}')} \mathscr{L}_{eff} = 0$ produces at most a trivial solution.
So, admit that $\mathscr{L}_{eff}$ depends on the metric and its derivatives. The 00 and 11 components of the EE are, for $(\bar{f}, \bar{g})$,

\begin{equation}
\label{D5}
\frac{\bar{g}' r + \bar{g} -1}{r^2} = \bar{T_0^0} \ ,
\end{equation}

\begin{equation}
\label{D6}
\frac{\frac{\bar{f}'}{\bar{f}} \bar{g} r + \bar{g} -1}{r^2} = \bar{T_1^1} \ ,
\end{equation}

while the pair $(f, g)$ produces a similar result. The first equation can be easily rewritten in the integral form (we work both for $\bar{g}$ and for $g$)

\begin{equation}
\label{D7}
\bar{g}(r) = 1 + \frac{a}{r} + \frac{1}{r} \int r^2 \bar{T_0^0} \ ,
\end{equation}

\begin{equation}
\label{D8}
g(r) = 1 + \frac{b}{r} + \frac{1}{r} \int r^2 T_0^0 \ ,
\end{equation}

where $a$ and $b$ are integration constants. We have that, whatever is the solution of eq. (\ref{D3})-(\ref{D4}), 

\begin{equation}
\label{D9}
g - \bar{g} = \frac{b-a}{r} + \frac{1}{r} \int r^2 \left( T_0^0 - \bar{T_0^0} \right) = \frac{b-a}{r} = \frac{c}{r} \ ,
\end{equation}

due to the independence on the solution of the 00 component of the SET.

Study now the second EE. The solution can be expressed via the following integral equations (once again, we work both on $\bar{f}$ and $f$):

\begin{equation}
\label{D10}
\bar{f} (r) = \frac{a'}{r} \exp \left( \int \frac{1}{\bar{g}r} \left( 1 + r^2 \bar{T_1^1} \right) \right) \ ,
\end{equation}

\begin{equation}
\label{D11}
f(r) = \frac{b'}{r} \exp \left( \int \frac{1}{gr} \left( 1 + r^2 T_1^1 \right) \right) \ ,
\end{equation}

where $a'$ and $b'$ are two integration constants. Calculating $\bar{f}-f$ from eq. (\ref{D6}) and its counterpart and from eq. (\ref{D10})-(\ref{D11}), $a'$ and $b'$ turn to be equal. Thus we have that

\begin{equation}
\label{D12}
\begin{split}
f - \bar{f} &= \frac{a'}{r} \left( \exp \left( \int \frac{1}{gr} \left( 1 + r^2 T_1^1 \right) \right) - \exp \left( \int \frac{1}{\bar{g}r} \left( 1 + r^2 \bar{T_1^1} \right) \right) \right) \\
&= \frac{a'}{r} \left( \frac{f}{g} \exp \left( \int \frac{1}{gr} \left( 1 + r^2 T_0^0 \right) \right) - \frac{\bar{f}}{\bar{g}} \exp \left( \int \frac{1}{\bar{g}r} \left( 1 + r^2 \bar{T_0^0} \right) \right) \right) \\
&= \frac{a'}{r} \left( \frac{f}{g} \left( b + \int \left( 1 + r^2 T_0^0 \right) \right) - \frac{\bar{f}}{\bar{g}} \left( a + \int \left( 1 + r^2 \bar{T_0^0} \right) \right) \right) \ .
\end{split}
\end{equation}

In the second line, we used the EE to simplify this object, while in the third line, we used eq. (\ref{D7})-(\ref{D8}). Following our assumption, $f - \bar{f} \propto 1/r$, meaning that the term in parentheses should be a constant; but this is impossible to be guaranteed in general, just from the simple fact that $a$ and $b$ are fully arbitrary and cannot be suitably tuned. It turns that the only way to make this object constant is to fix $T_0^0 = T_1^1 = 0$, but this means that $(f, g)$ reduces to the Schwarzschild solution and this happens only in vacuo (that is, $\mathscr{L}_{eff} = 0$). It turns out that, in order to satisfy the relation among $(f, g)$ and $(\bar{f}, \bar{g})$, there is a single possibility, i.e. $\partial_{(\bar{f}, \bar{g})} \mathscr{L}_{eff} = 0$ and $\partial_{(\bar{f'}, \bar{g'})} \mathscr{L}_{eff} = 0$. This closes the proof.

Notice that we should require both conditions $\partial_{(\bar{f}, \bar{g})} \mathscr{L}_{eff} = 0$ and $\partial_{(\bar{f'}, \bar{g'})} \mathscr{L}_{eff} = 0$ to hold. Indeed, if we limited ourselves to a single one, we still would have a non-vanishing SET in eq. (\ref{D12}), so that the whole previous argument holds.

En passant, we are also able to prove that (a) $\Rightarrow$ (c). Indeed, since we proved that (a) $\Rightarrow$ (b), this applies on the SET as

\begin{equation}
\label{D13}
T_0^0 = - \frac{1}{2} \mathscr{L}_{eff} = T_1^1 \ ,
\end{equation}

and this is a renowend necessary and sufficient condition for having $f=g$. Notice that this does not imply the equivalence among (a) and (c), since this also requires (c) $\Rightarrow$ (a).

\subsection{ $(f, g) \ \text{is a solution} \qquad \Leftarrow \qquad \partial_{(\bar{f}, \bar{g})} \mathscr{L} = 0 \ \text{and} \ \partial_{(\bar{f}', \bar{g}')} \mathscr{L} = 0$}

This second proof turns to be quite simpler, since we only need to proceed with a direct calculation. Indeed, the SET results to be

\begin{equation}
\label{D13}
\begin{split}
T_0^0 &= - f \partial_f \mathscr{L}_{eff} - \frac{1}{2} \mathscr{L}_{eff} + 2f' \partial_{f'} \mathscr{L}_{eff} + f \frac{\partial}{\partial r} \partial_{f'} \mathscr{L}_{eff} \\
&= - \frac{1}{2} \mathscr{L}_{eff}
\end{split}
\end{equation}

\begin{equation}
\label{D14}
\begin{split}
T_1^1 &= f \partial_f \mathscr{L}_{eff} - \frac{1}{2} \mathscr{L}_{eff} - f \frac{\partial}{\partial r} \partial_{f'} \mathscr{L}_{eff} \\
&= - \frac{1}{2} \mathscr{L}_{eff} \ .
\end{split}
\end{equation}

The resulting condition $T_0^0 = T_1^1$ is enough to state that $f=g$. This allows to write the independent component of the Einstein tensor as

\begin{equation}
\label{D15}
G_0^0 = \frac{f'r + f - 1}{r^2} = \frac{\bar{f}' r - \frac{c}{r^2} r + \bar{f} + \frac{c}{r} - 1}{r^2} = \frac{\bar{f}' r + \bar{f} - 1}{r^2} = \bar{G_0^0} \ ,
\end{equation}

\begin{equation}
\label{D16}
G_2^2 = \frac{f''r + 2f'}{2r} = \frac{\bar{f}'' r + \frac{2c}{r^3} r + 2\bar{f}' - 2\frac{c}{r^2} - 1}{2r} = \frac{\bar{f}'' r + 2\bar{f}' - 1}{2r} = \bar{G_2^2} \ .
\end{equation}

On the other hand, we already proved that the SET components satisfy the related equivalences $T_0^0 = \bar{T_0^0}$ and $T_2^2 = \bar{T_2^2}$: the proof is immediate from eq. (\ref{D13})-(\ref{D15}). Thus, since equal things are equal, the EE hold also for $(f, g)$ and this closes the proof.

Before closing the section, we stress that we did not prove the implication (c) $\Rightarrow$ (a), but only that (b) $\Rightarrow$ (c) $\Rightarrow$ (a); and this is not the same thing.

\section{Proof of the theorem $-$ part 2}

In the following, we will prove statement (\ref{B8}). The equivalence among (a) and (b) already holds, because nothing changes from the previous proof (eventually, things are even simpler). Here we only need to prove that also (c) is equivalent to (a) and (b). Thus we perform the extra assumption, i.e. $\partial_{(\bar{f}, \bar{g})} \mathscr{L} = 0$ or $\partial_{(\bar{f}', \bar{g}')} \mathscr{L} = 0$. The implication $\Rightarrow$ has already been proved, so we only need to check the opposite arrow; we will do it in both cases, first assuming $\partial_{(\bar{f}, \bar{g})} \mathscr{L} = 0$ and then $\partial_{(\bar{f}', \bar{g}')} \mathscr{L} = 0$.

\subsection{$(f, g)$ is a solution \qquad $\Leftarrow$ \qquad $\bar{f} = \bar{g}$ and $\partial_f \mathscr{L}_{eff} = 0$}

We assume that the effective Lagrangian does not depend on the metric, but still may depend on its derivatives. We only need to prove that the Einstein tensor and the SET are equal in both the cases of $(f, g)$ and $(\bar{f}, \bar{g})$.

The first part is easy. Indeed, the independent components of the EE are

\begin{equation}
\label{E1}
G_0^0 = \frac{f'r + f - 1}{r^2} = \frac{\bar{f}'r - \frac{c}{r^2}r + \bar{f} + \frac{c}{r} - 1}{r^2} = \frac{\bar{f}'r + \bar{f} - 1}{r^2} = \bar{G_0^0} \ ,
\end{equation}

\begin{equation}
\label{E2}
G_2^2 = \frac{f''r + 2f'}{2r} = \frac{\bar{f}''r - \frac{2c}{r^3}r + 2\bar{f}' + 2\frac{c}{r^2}}{2r} = \frac{\bar{f}''r + 2\bar{f}'}{2r} = \bar{G_2^2} \ .
\end{equation}

We are left to prove that also the SET behaves in the same way. Now, since we assumed that $\bar{f}=\bar{g}$, a consequence of it is that $\bar{T_0^0} = \bar{T_1^1}$. However, since this is also a necessary and sufficient condition for it, if we want to have $f=g$, the same must hold also for the unbarred components with the unbarred functions, i.e. $T_0^0 = T_1^1$; otherwise, there is no hope for $f=g$ to be a solution.

This means aking that

\begin{equation}
\label{E3}
- f \partial_f \mathscr{L}_{eff} - \frac{1}{2} \mathscr{L}_{eff} + 2f' \partial_{f'} \mathscr{L}_{eff} + f \frac{\partial}{\partial r} \partial_{f'} \mathscr{L}_{eff} = f \partial_f \mathscr{L}_{eff} - \frac{1}{2} \mathscr{L}_{eff} - f \frac{\partial}{\partial r} \partial_{f'} \mathscr{L}_{eff} \ ,
\end{equation}

that is

\begin{equation}
\label{E4}
f \partial_f \mathscr{L}_{eff} - \frac{\partial}{\partial r} \left( f \partial_{f'} \mathscr{L}_{eff} \right) = 0 \ .
\end{equation}

This is a general condition, holding for any Lagrangian. However, since in this subsection we are assuming $\partial_f \mathscr{L}_{eff} = 0$, eq. (\ref{E4}) reduces to

\begin{equation}
\label{E5}
- \frac{\partial}{\partial r} \left( f f \partial_{f'} \mathscr{L}_{eff} \right) = 0 \qquad \Rightarrow \qquad f' \partial_f \left( f \partial_{f'} \mathscr{L}_{eff} \right) = f' \partial_{f'} \mathscr{L}_{eff} = 0 \ ,
\end{equation}

which is clearly false, unless $\partial_{f'} \mathscr{L}_{eff} = 0$. This means that (c) $\Rightarrow$ (b) and, due to the equivalence among (a) and (b), this is enough to close the proof.

\subsection{$(f, g)$ is a solution \qquad $\Leftarrow$ \qquad $\bar{f} = \bar{g}$ and $\partial_{f'} \mathscr{L}_{eff} = 0$}

This proof is even simpler. Since it rests only on general assumptions, the discussion along eq. (\ref{E1})-(\ref{E4}) is exactly the same also in this case. This time, if $\partial_{f'} \mathscr{L}_{eff} = 0$, eq. (\ref{E4}) becomes

\begin{equation}
\label{E6}
f \partial_f \mathscr{L}_{eff} = 0 \ .
\end{equation}

One immediately sees that also $\partial_f \mathscr{L}_{eff} = 0$, thus having (c) $\Rightarrow$ (b); the equivalence among (a) and (b) closes the proof.

\section{Corollary}

In \cite{Chinaglia18} we already proved, as a corollary, that the only non-Lagrangian singularity of the solution has a Schwarzschild-like behaviour $\propto 1/r$. This actually holds also in the more general version of the theorem, both assuming statement (\ref{B7}) and statement (\ref{B8}). Since all the assumptions performed for (\ref{B7}) also appear in (\ref{B8}), it is enough to prove the corollary for (\ref{B7}); the proof for (\ref{B8}) will consequently follow.

The statement of the corollary is the following: 

\bigskip

\begin{minipage}{.8\textwidth}

{\it Holding the theorem, the only non-Lagrangian singularity is Schwarzschild-like.}

\begin{equation}
\label{F1}
(\bar{f}+h, \bar{g}+h) \ \text{is a solution} \qquad \Rightarrow \qquad h = \frac{c}{r} \ ,
\end{equation}

\end{minipage}

\bigskip

where $c$ is an integration constant. Just for simplicity, we rewrite $h$ as $h = \tilde{h} + c/r$, isolating a Schwarzschild term so that $\tilde{h}$ does not contain any term proportional to $1/r$. Thus the statement reduces to $\tilde{h}=0$.

Before proceeding, we note that here we don't deal with those singularities coming from the matter Lagrangian. As example, in the Reissner-Nordström solution, a singular term appears, proportional to $1/r^2$; that term, however, directly arises due to the matter term in the Lagrangian (and indeed can be cured, just smearing the Maxwell term, in the spirit of Born-Infeld approach \cite{Born-Infeld}).

\subsection{$(\bar{f}+h, \bar{g}+h) \ \text{is a solution} \qquad \Rightarrow \qquad \tilde{h}=0$}

We start writing the independent components of the EE. Since we are assuming the theorem to hold, we are supposing (a) and (b) to be verified, but we already proved that (a), (b) $\Rightarrow$ (c); so that the only two independent components of the EE are

\begin{equation}
\label{F2}
\frac{\bar{f}'r + \bar{f} - 1}{r^2} = \bar{T_0^0} \ ,
\end{equation}

\begin{equation}
\label{F3}
\frac{\bar{f}''r + 2 \bar{f}'}{2r} = \bar{T_2^2} \ ,
\end{equation}

with respect to $\bar{f}$. On the other hand, with respect to $\bar{f}+h$ we have

\begin{equation}
\label{F4}
\frac{\bar{f}'r + \tilde{h}r - \frac{c}{r^2} r + \bar{f} + \tilde{h} + \frac{c}{r} - 1}{r^2} = \frac{\bar{f}'r + \bar{f} - 1}{r^2} + \frac{\tilde{h}r + \tilde{h}}{r^2} = \bar{T_0^0} |_h \ ,
\end{equation}

\begin{equation}
\label{F5}
\frac{\bar{f}''r + \tilde{h}''r + \frac{2c}{r^3} r + 2 \bar{f}' + 2\tilde{h}' - 2 \frac{c}{r^2}}{2r} = \frac{\bar{f}''r + 2 \bar{f}'}{2r} + \frac{\tilde{h}''r + 2\tilde{h}'}{2r} = \bar{T_2^2} |_h \ ,
\end{equation}

where the subscript $|_h$ indicates that the tensor is calculated for $\bar{f}+h$. However, since statement (b) holds,  the SET does not depend on the metric, so that $\bar{T_{\mu}^{\nu}} |_h = \bar{T_{\mu}^{\nu}}$. And since we are assuming $\bar{f}+h$ to be a solution too, this immediately implies $\tilde{h}r + \tilde{h} = 0$ and $\tilde{h}''r + 2\tilde{h}' = 0$, conditions which can be satisfied only by a null or a Schwarzschild-like solution; however, since we already put a Schwarzschild-like term in evidence from $h$, the only possible conclusion is $\tilde{h}=0$.

\subsection{$(\bar{f}+h, \bar{g}+h) \ \text{is a solution} \qquad \Leftarrow \qquad \tilde{h}=0$}

In this case, the proof is trivial: indeed, since $\tilde{h}=0$, $h$ reduces to a Schwarzschild-like term, i.e. $h = \tilde{h} + c/r = c/r$. But, since the theorem holds, $(\bar{f}+c/r, \bar{g}+c/r)$ actually is a solution.

\section{Comments}

In this paper, we discussed and proved a more general version of the theorem already presented in \cite{Chinaglia18}. At the time, we argued that the proof could not be complete, mainly due to the restriction on the Lagrangian: indeed, accepting a non minimal coupling, there are at least two counterexamples of the theorem. The first is the already mentioned BLZ black hole \cite{Balakin16, Balakin16_1}, while the second is provided by the so called Non-Polynomial Gravity (NPG) approach (a mention is also in \cite{Chinaglia18}, while more detailed reviews can be found in \cite{Colleaux17, Colleaux18}). In both cases, statement (c) was satisfied, while statements (a) and (b) were not. By the way, these two counterexamples prove that the extra assumpation which distinguishes statement (\ref{B7}) from statemente (\ref{B8}) are actually not only sufficient, but also necessary.

This new theorem takes into account any possible Lagrangian, while in the original case we only discussed the minimally coupled one (such as the NED model we brought as example). As we should expect, the result is much more limited, especially because statement (c) needs a very specific condition to be equivalent to (a) and (b). This allows us to build up a large number of theories for which $f=g$, but without any Schwarzschild term to come into account.

The theorem, indeed, gives us a criterion to establish if the (effective) Lagrangian will present or not a singular $c/r$ term. If we are not interested on the $g_{00}g_{11} = -1$ condition, it is enough to violate statement (b), i.e. asking $\mathscr{L}_{eff}$ to depend either to the solution or to its derivative. As example of this, consider a scalar field minimally coupled to gravity: in this case, $- \mathscr{L}_{eff} = \frac{1}{2} \nabla_{\mu} \phi \nabla^{\mu} \phi - V(\phi) = \frac{1}{2} g {\phi'}^2 - V(\phi)$. Since $\phi$ is a scalar, $V(\phi)$ depends neither on the metric, nor on its derivatives; the presence of $g$ inside the kinetic term guarantees that the term $c/r$ will not appear. Of course, as expicitly shown in \cite{Chinaglia18}, this implies $f \neq g$.

In \cite{Chinaglia18} we also proved that the theorem holds in any dimension, while the corollary holds only for $D \geq 4$. Here we gave the proof only in the 4D case, but the general argument can be easily transposed in more (or less) dimensions, so that we exepect the theorem to hold also in that case. This possibility may be discussed later on in a future addendum.

Finally, if we want to have $f=g$ but we don't want to deal with any singualrity, the Lagrangian must be provided by the metric {\it and} its derivatives (as we already proved in this paper): otherwise, there is no hope to satisfy both statement (c) and eq. (\ref{E4}).

A brief note: in \cite{Balakin16}, the effective SET depends only on the derivative of the metric, but not on the metric itself. However, this does not violate our criterion, since the presence of the solution into the Lagrangian does not imply its presence into the SET too (as we already argued in the Lemma). As example, the effective Lagrangian $\mathscr{L}_{eff} = 2A + \frac{2}{3}B'f + Bf$, where $A$ and $B$ and do not depend on the metric or its derivatives and such that $f=g$, generates the SET components $T_0^0 = T_1^1 = - A + \frac{3}{2} Bf'$, independent on the metric itself.

\end{document}